\documentclass[10pt]{article}
\usepackage{geometry}
\usepackage{xcolor}
\definecolor{graycolor}{gray}{0.9} 
\usepackage{microtype}
\usepackage{setspace} 
\usepackage[utf8]{inputenc}
\usepackage[english]{babel}
\usepackage{times}
\usepackage{array}
\usepackage{soul}
\usepackage{cite}
\usepackage[numbers,sort&compress]{natbib}
\usepackage{natbib}

\usepackage{titlesec} 
\titleformat {\section} [block] {\raggedright \fontsize{10}{10}\selectfont\bfseries} {\thesection. \space} {0pt} {}
\titlespacing {\section} {0pt} {12pt} {6pt}
\titleformat {\subsection} [block] {\raggedright \fontsize{10}{10}\selectfont\itshape} {\thesubsection .\space} {0pt} {}
\titlespacing {\subsection} {0pt} {12pt} {6pt}
\titleformat {\subsubsection} [block] {\raggedright \fontsize{10}{10}\selectfont} {\thesubsubsection .\space} {0pt} {}
\titlespacing {\subsubsection} {0pt} {12pt} {6pt}
\titleformat {\paragraph} [block] {\raggedright \fontsize{10}{10}\selectfont} {} {0pt} {}
\titlespacing {\paragraph} {0pt} {12pt} {6pt}

\usepackage{array} \newcommand{\PreserveBackslash}[1]{\let\temp=\\#1\let\\=\temp}
\newcolumntype{C}[1]{>{\PreserveBackslash\centering}m{#1}}
\newcolumntype{R}[1]{>{\PreserveBackslash\raggedleft}m{#1}}
\newcolumntype{L}[1]{>{\PreserveBackslash\raggedright}m{#1}}
\usepackage{lineno}
\usepackage{tabularx}
\usepackage{colortbl}
\usepackage{graphicx}
\usepackage{float}
\usepackage[export]{adjustbox}
\usepackage{caption}
\usepackage{fancyhdr} 
\usepackage{amsmath}
\usepackage{ upgreek }
\usepackage{graphicx}
\usepackage{cmap}
\usepackage[utf8]{inputenc}
\usepackage[T1]{fontenc}
\usepackage{amsfonts}
\usepackage{ amssymb }
\usepackage{lastpage}
\usepackage{layout}
\usepackage{setspace} 
\usepackage{enumitem}
\usepackage{booktabs}
\usepackage{arydshln}
\usepackage{multirow}
\usepackage{color}
\usepackage{hyperref} 
\hypersetup{
	colorlinks=true,
	linkcolor=blue,
	filecolor=blue,
	urlcolor=black,
	citecolor=cyan,
}

\soulregister\cite7
\soulregister\ref7
\setcitestyle{open={[},close={]},citesep={,\!},numbers}

\fancypagestyle{firstpage}{
    \setlength{\headsep}{2.2cm}
    
    \setlength{\footskip}{1.5cm}
    \fancyhf{}
    \lhead{\begin{table}[H]
        \centering
        \begin{tabular}{L{2.5cm}C{10cm}C{3.1cm}R{2cm}}
            \includegraphics[scale=0.035]{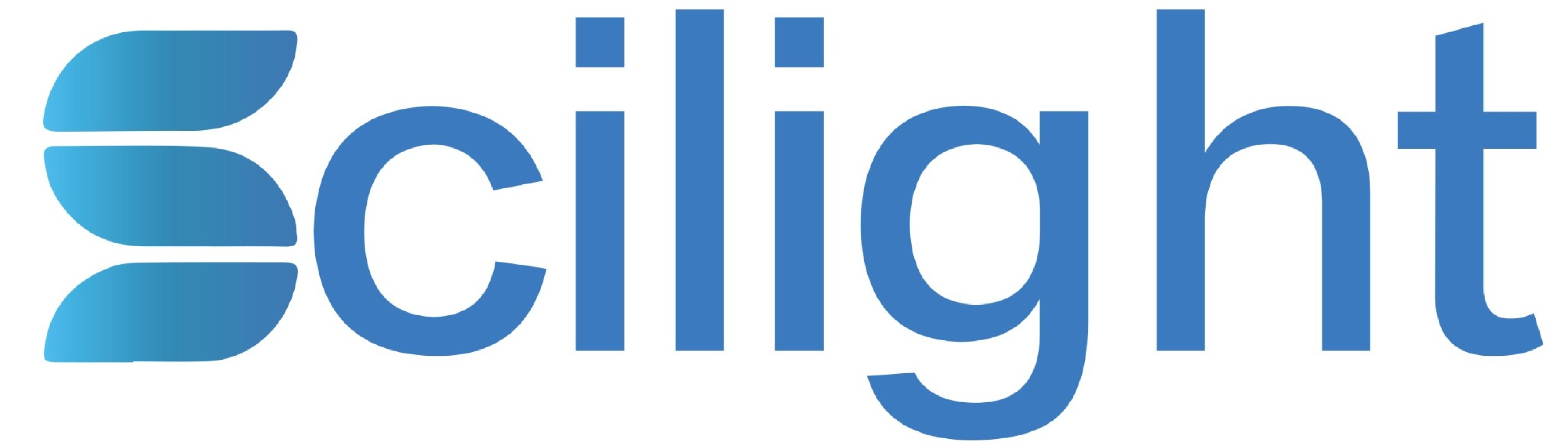} \vspace{-6pt}& \cellcolor{graycolor}\begin{tabular}[c]{@{}c@{}}\textit{International Journal of Gravitation and Theoretical Physics}\\ \href{https://www.sciltp.com/journals/ijgtp}{https://www.sciltp.com/journals/ijgtp}\end{tabular} & \includegraphics[scale=0.020]{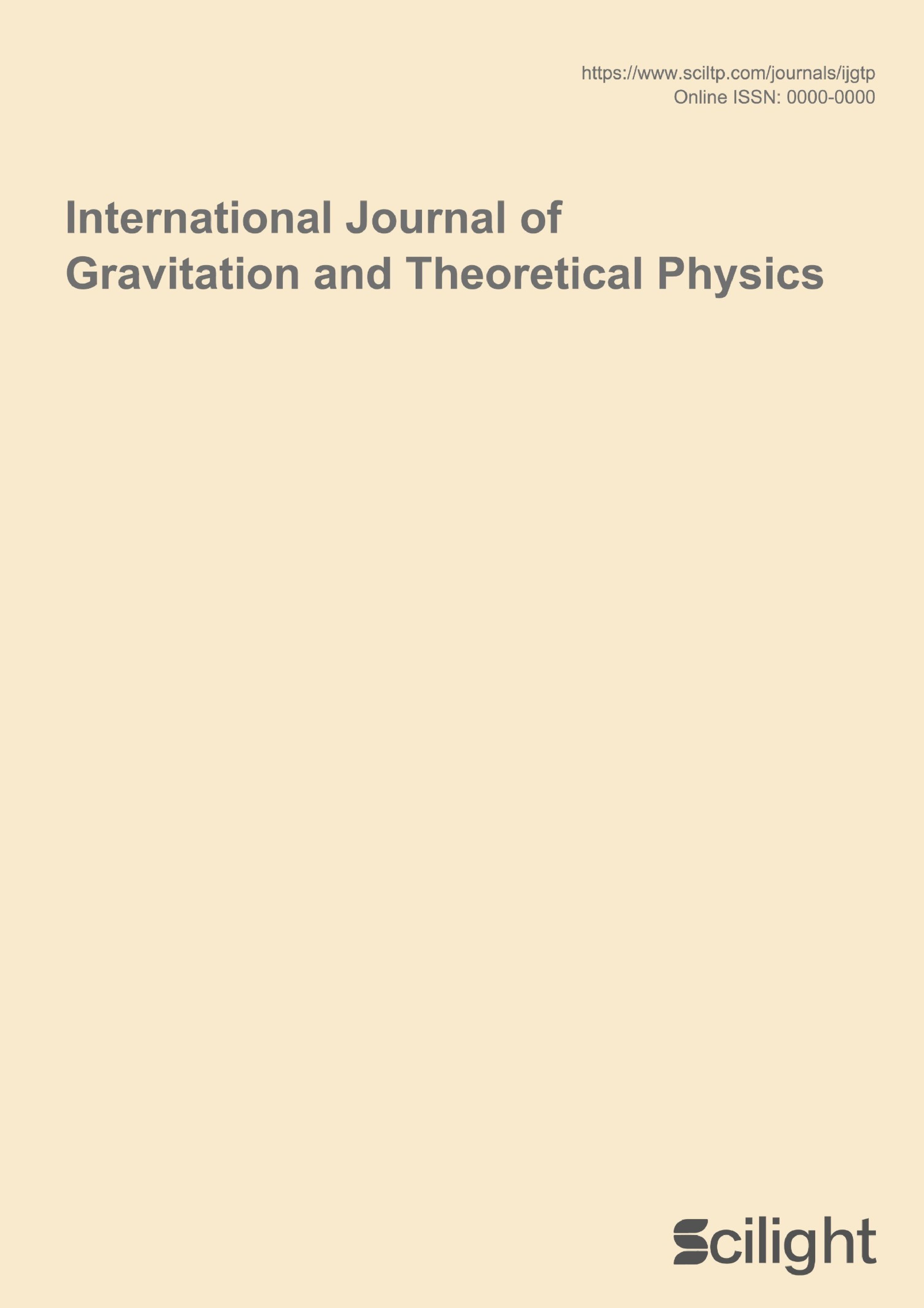} \vspace{-3pt}\\
        \end{tabular}
        \vspace{-22pt}
    \end{table}}
   
    \fancyfoot[C]{
        \vspace{-1.55cm}
        \begin{table}[H]
            \begin{minipage}[c]{0.15\columnwidth}
                \includegraphics[scale=0.04]{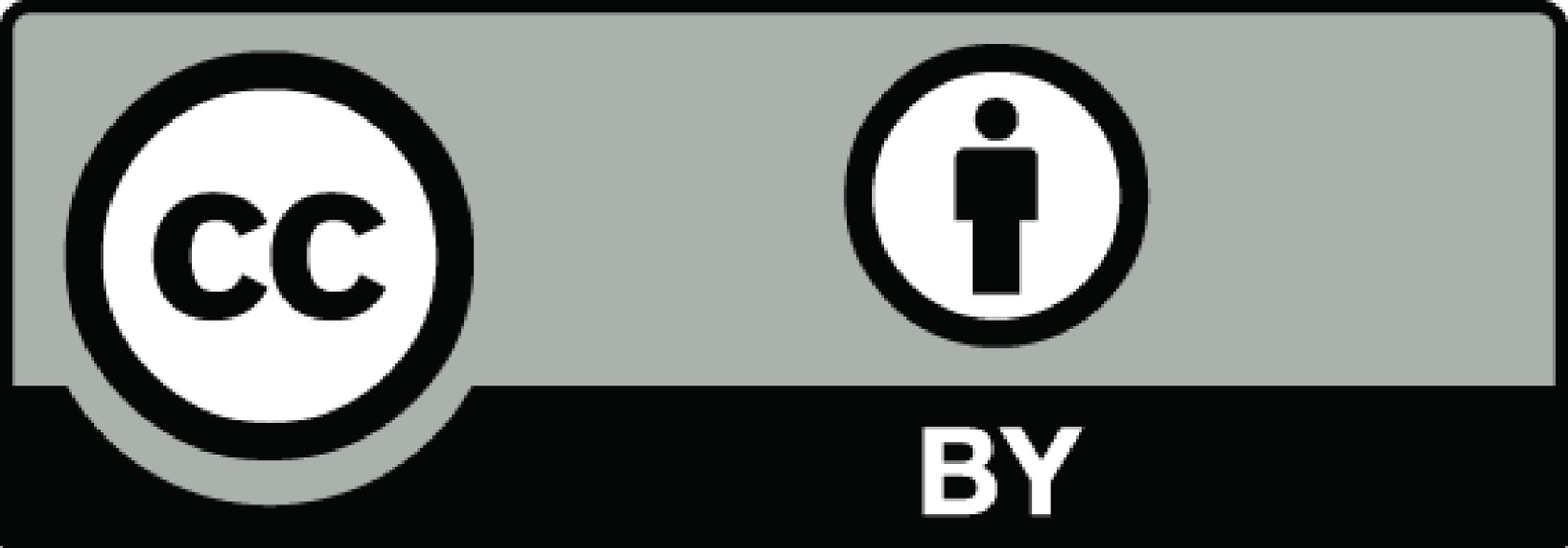} \vspace{1.1pt}
            \end{minipage}
            \hfill
            \begin{minipage}[c]{0.85\columnwidth}
                \scriptsize \textbf{Copyright:} © 2025 by the authors. This is an open access article under the terms and conditions of the Creative Commons Attribution (\mbox{CC BY}) license (\href{https://creativecommons.org/licenses/by/4.0/}{https://creativecommons.org/licenses/by/4.0/}). \\ \textbf{Publisher’s Note:} Scilight stays neutral with regard to jurisdictional claims in published maps and institutional affiliations.
            \end{minipage}
    \end{table}}
    \vspace{-0.55cm}
}

\begin{document}
\newgeometry{left=2.5cm, right=2.5cm, top=1.8cm, bottom=4cm}
	\thispagestyle{firstpage}
	\nolinenumbers
	{\noindent \textit{Article} 
}
	\vspace{4pt} \\
	{\fontsize{18pt}{10pt}\textbf{Black Holes in Proca-Gauss-Bonnet Gravity with Primary Hair:  Particle Motion, Shadows, and Grey-Body Factors}  }
	\vspace{16pt} \\
	{\large Bekir Can 
 Lütfüoğlu }
	\vspace{6pt}
	 \begin{spacing}{0.9}
		{\noindent \small
				Department of Physics, Faculty of Science, University of Hradec Kralove, Rokitanskeho 62/26, 500 03 Hradec Kralove, \mbox{Czech Republic}; bekir.lutfuoglu@uhk.cz \vspace{6pt}\\
		\footnotesize	\textbf{How To Cite}:Lütfüoğlu, B.C. Black Holes in Proca-Gauss-Bonnet Gravity with Primary Hair:  Particle Motion, Shadows, and Grey-Body Factors. \emph{International Journal of Gravitation and Theoretical Physics} \textbf{2025}, \emph{1}(1), 4. }\\
	\end{spacing}

\begin{table}[H]
\noindent\rule[0.15\baselineskip]{\textwidth}{0.5pt} 
\begin{tabular}{lp{12cm}}  
 \small 
  \begin{tabular}[t]{@{}l@{}} 
  \footnotesize  Received: day month year \\
  \footnotesize  Revised: day month year \\
   \footnotesize Accepted: day month year \\
  \footnotesize  Published: day month year
  \end{tabular} &
  \textbf{Abstract:} We investigate classical and semiclassical signatures of black holes in a recently proposed Proca–Gauss–Bonnet gravity model that admits asymptotically flat solutions with primary hair. Two distinct classes of spherically symmetric metrics arise from different relations between the coupling constants of scalar–tensor and vector–tensor Gauss–Bonnet interactions. For each geometry, we examine the range of parameters permitting horizon formation and analyze the motion of test particles and light rays. We compute characteristic observables including the shadow radius, Lyapunov exponent, innermost stable circular orbit (ISCO) frequency, and binding energy. Additionally, we study scalar and Dirac field perturbations, derive the corresponding effective potentials, and calculate the grey-body factors (GBFs) using both the sixth-order Wentzel–Kramers–Brillouin (WKB) method and their correspondence with quasinormal modes (QNMs). Our results show that the QNM-based approximation of GBFs is accurate for sufficiently large multipole numbers and that deviations from Schwarzschild geometry become pronounced for large values of the Proca hair and Gauss–Bonnet couplings. \\
\\
  & 
  \textbf{Keywords:} black holes with primary hair; modified gravity; grey-body factors
\end{tabular}
\noindent\rule[0.15\baselineskip]{\textwidth}{0.5pt} 
\end{table}

	\section{Introduction }
	Black holes provide a unique arena for testing gravity in the strong-field regime. With the advent of observational tools such as the Event Horizon Telescope~\cite{EventHorizonTelescope:2019dse} and gravitational wave observatories like LIGO/Virgo~\cite{LIGOScientific:2016aoc}, theoretical models of black holes can now be scrutinized by direct comparison with empirical data \cite{EventHorizonTelescope:2022xqj,LIGOScientific:2016lio,Bambi:2015kza,Goddi:2016qax}. Among the most compelling targets for such tests are alternative or quantum-corrected black hole geometries that deviate from the classical predictions of general relativity, especially near the horizon.

A wide range of such geometries has been proposed in the literature, often inspired by extensions of Einstein's theory involving higher-curvature corrections, non-minimal couplings, or effective field theoretic completions. A particularly appealing class of models includes those that allow for analytically tractable, spherically symmetric solutions incorporating scalar and vector fields. These frameworks not only admit well-defined black hole metrics but also encode corrections expected from semiclassical or quantum gravitational considerations.

In a recent work~\cite{Charmousis:2025jpx}, a new family of black hole solutions was constructed by combining scalar–tensor and vector–tensor Gauss–Bonnet-type interactions into a unified framework. The resulting theory is governed by the action ~\cite{Charmousis:2025jpx}
\begin{equation}\label{action}
S = \int d^4x \sqrt{-g} \left( R - \alpha\, \mathcal{L}^{\text{VT}}_G - \beta\, \mathcal{L}^{\text{ST}}_G \right),
\end{equation}
where \( \mathcal{L}^{\text{VT}}_G \) and \( \mathcal{L}^{\text{ST}}_G \) denote vector–tensor and scalar–tensor Gauss–Bonnet-like Lagrangians, and \( \alpha \), \( \beta \) are coupling constants. Remarkably, this setup admits exact, asymptotically flat black hole solutions that exhibit \textit{primary hair}, i.e., independent integration constants or a global charge, which is not tied to other conserved charges like mass or electric charge.

\textls[15]{Two distinct metric functions emerge in \cite{Charmousis:2025jpx} depending on the relation between \( \alpha \) and \( \beta \). For generic values}
\restoregeometry

\noindent \( \beta \ne -\alpha \), the solution involves a square-root structure characteristic of non-local corrections. In the special case \( \beta = -\alpha \), the solution simplifies to a rational form with regular behavior for a suitable parameter range. In both cases, the metrics reduce to the Schwarzschild solution in a particular limit. These properties make the two geometries particularly attractive for probing both classical and quantum aspects of black hole physics.

In this paper, we undertake a detailed analysis of classical and semiclassical observables in the background of these two black hole solutions. On the classical side, we investigate the motion of test particles and light rays, focusing on the effective potential, circular geodesics, orbital velocity at the ISCO, and binding energy. We compute the radius of the photon sphere and black hole shadow, along with the associated Lyapunov exponent that governs the stability of circular photon orbits. On the semiclassical side, we study the propagation of test scalar and spinor fields, evaluate the GBFs, and discuss their implications for Hawking radiation spectra.

The structure of the paper is as follows. In Section~\ref{sec:metrics}, we review the two static, spherically symmetric black hole solutions arising in Proca–Gauss–Bonnet gravity and analyze the parameter regimes admitting event horizons. Section~\ref{sec:particleshadows} is devoted to the study of particle motion and black hole shadows, where we extract key observables such as the photon sphere radius, Lyapunov exponent, ISCO frequency, and binding energy. In Section~\ref{sec:GBF}, we analyze scalar and spinor field perturbations, derive the associated effective potentials, and compute GBFs using both the higher-order WKB method and their correspondence with QNMs. Finally, Section~\ref{sec:conclusion} summarizes our results and discusses their implications for tests of black hole physics beyond general relativity.

\section{Black Hole Metrics and Basic Properties}\label{sec:metrics}

We consider two static, spherically symmetric solutions recently proposed in \cite{Charmousis:2025jpx}, given by the following line element:
\begin{equation}
ds^2 = -f(r)\, dt^2 + \frac{dr^2}{f(r)} + r^2 (d\theta^2 + \sin^2 \theta\, d\phi^2),
\end{equation}
where, for $\alpha\neq\beta$, the metric function \( f(r) \) is
\begin{align}\label{metricfunc}
f(r) &= 1 - \frac{2 \alpha (M - Q)}{r (\alpha + \beta)}  + \frac{r^2}{2 (\alpha + \beta)} \\\nonumber
& - \frac{r^2}{2 (\alpha + \beta)}   
\sqrt{1 + \frac{8 \alpha Q}{r^3} + \frac{8 \beta M}{r^3} + \frac{16 \alpha \beta (M - Q)^2}{r^6} },
\end{align}
where \( M \) and \( Q \) are integration constants associated with mass and an additional charge-like parameter, while \( \alpha \) and \( \beta \) are the coupling parameters, respectively, for the vector–tensor and scalar–tensor corrections in (\ref{action}).
The charge $Q$, associated with the Proca field,  cannot be expressed in terms of mass $M$ and, thereby, represents the primary hair. When $Q=M$ we reproduce the black-hole metric in the scalar-tensor version of the 4D Einstein-Gauss-Bonnet theory \cite{Lu:2020iav,Kobayashi:2020wqy,Fernandes:2020nbq} 

In the limit $\beta\to -\alpha$, the solution (\ref{metricfunc}) takes the following form:
\begin{align}
\lim_{\beta\to-\alpha}f(r) &= \frac{r^3}{r^3 - 4 \alpha (M - Q)} \Bigg( 1 - \frac{2 M}{r} + \frac{4 \alpha (M - Q)^2}{r^4} 
- \frac{4 \alpha (M - Q)}{r^3} \Bigg).
\end{align}
Setting \( \alpha = \beta = 0 \)  yields the Schwarzschild solution.

For both metrics, the location of the event horizon \( r_h \) is determined by solving \( f(r_h) = 0 \). The existence and number of horizons depend on the values of \( \alpha \), \( \beta \), \( M \), and \( Q \). As \( r \to \infty \), both metrics asymptotically approach the Schwarzschild form,
\[
f(r) = 1 - \frac{2 M}{r} + \mathcal{O}({r^{-2}}),
\]
so that \( M \) is the Arnowitt-Deser-Misner (ADM) mass \cite{Arnowitt:1960es}.

\begin{figure}[H]
\centering
\resizebox{0.98\linewidth}{!}{\includegraphics{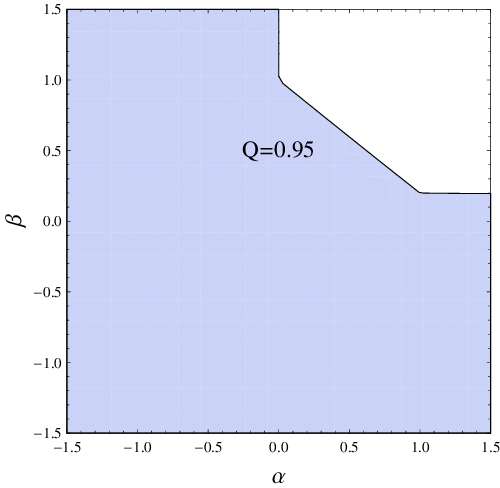}\includegraphics{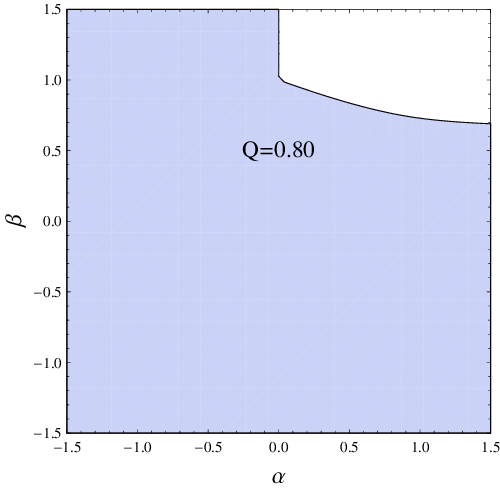}\includegraphics{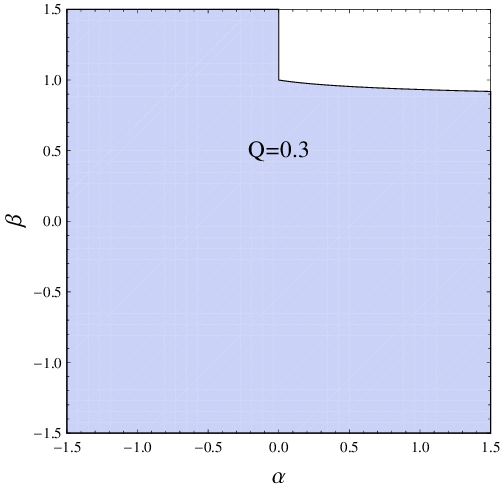}}
\resizebox{0.98\linewidth}{!}{\includegraphics{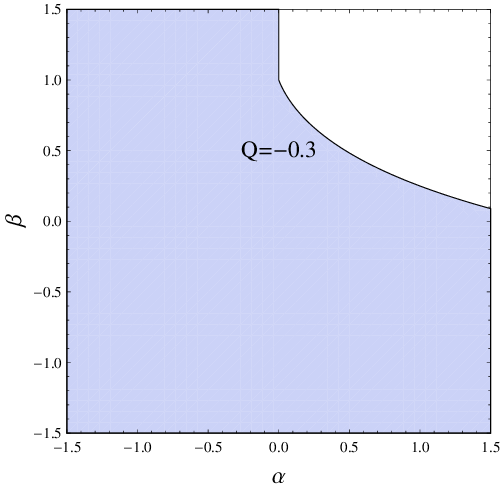}\includegraphics{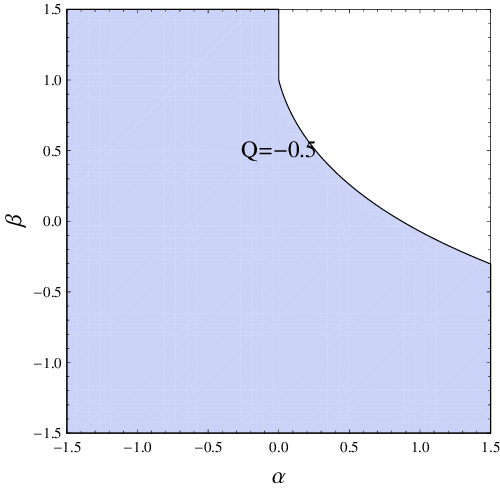}\includegraphics{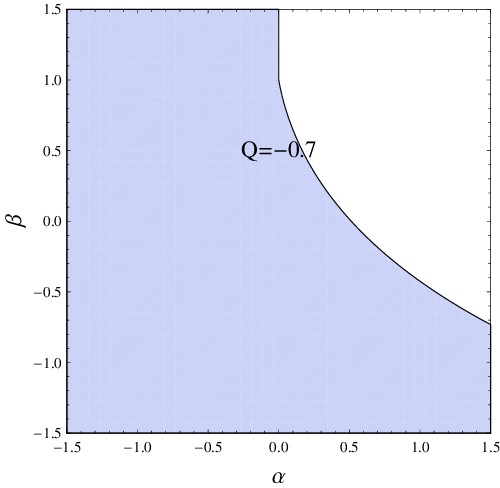}}
\caption{The range of parameters allowing for the existence of the event horizon, $Q=0.95$ (\textbf{top left}), $Q=0.8$ (\textbf{top middle}), $Q=0.3$ (\textbf{top right}), $Q=-0.3$ (\textbf{bottom left}), $Q=-0.5$ (\textbf{bottom middle}), $Q=-0.7$ (\textbf{bottom right}); $M=1$.}
\label{fig:BH1Range}
\end{figure}

\begin{figure}[H]
\centering
\resizebox{0.5\linewidth}{!}{\includegraphics{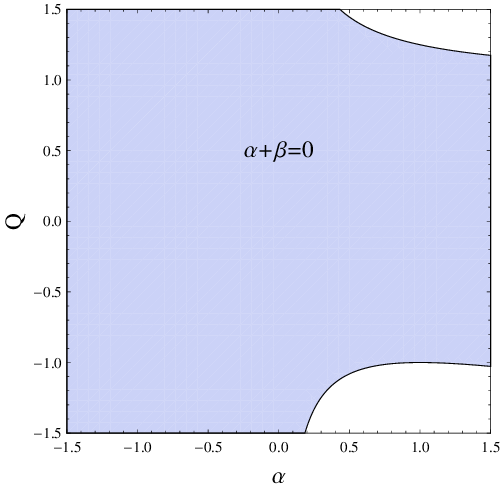}}
\caption{The range of parameters for $\alpha+\beta=0$ allowing for the existence of the event horizon, $M=1$.}
\label{fig:BH2Range}
\end{figure}

The parameter ranges that permit the existence of an event horizon are shown in Figures~\ref{fig:BH1Range} and \ref{fig:BH2Range}. As can be seen, sufficiently large positive values of $\alpha$ and $\beta$ do not support black hole solutions when $\alpha + \beta \neq 0$, and similarly, large positive values of $Q$ also prevent the formation of an event horizon when $\alpha + \beta = 0$ for either large positive or negative $\alpha$ ($\beta$).

In the following sections, we explore the impact of these modifications on classical observables such as particle orbits and shadows, as well as on scattering characteristics such as GBFs.

\vspace{-24pt}

\section{Particle Motion and Shadows}\label{sec:particleshadows}

In this section, we analyze four specific characteristics of particle motion: the radius of the shadow~\cite{Synge:1966okc}, the Lyapunov exponent~\cite{Cornish:2003ig} associated with the instability of the photon sphere, the orbital velocity at the ISCO, and the binding energy.

\begin{figure}[H]
\centering
\resizebox{0.99\linewidth}{!}{\includegraphics{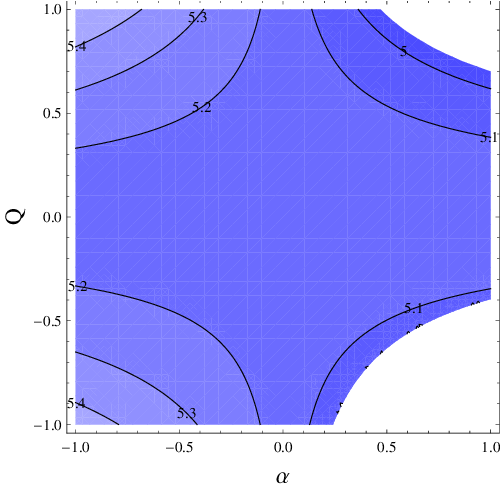}~~\includegraphics{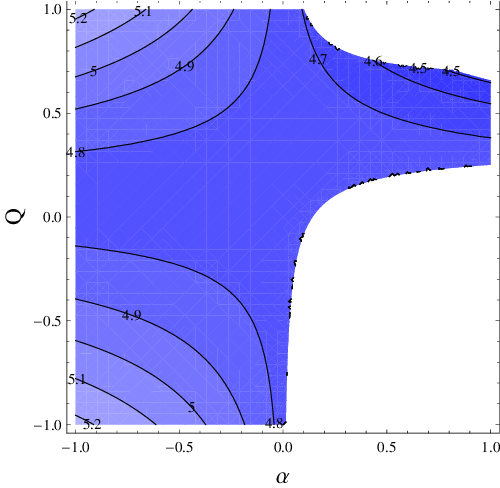}~~ \includegraphics{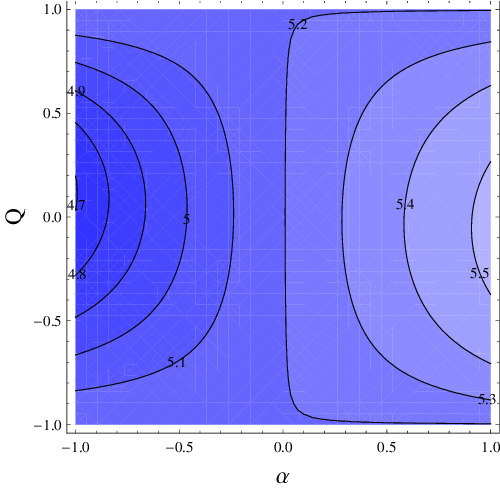}}
\caption{The radius of the shadow for $\beta=0.1$ (\textbf{left}), $\beta=0.9$ (\textbf{middle}) and $\alpha+\beta=0$ (\textbf{right}), $M=1$.}
\label{fig:Shadow}
\end{figure}

\begin{figure}[H]
\centering
\resizebox{0.98\linewidth}{!}{\includegraphics{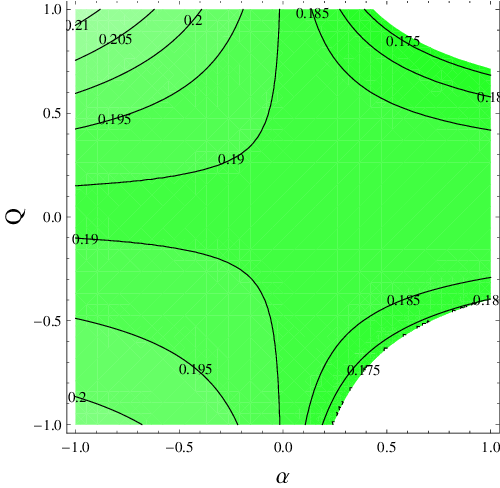}~~\includegraphics{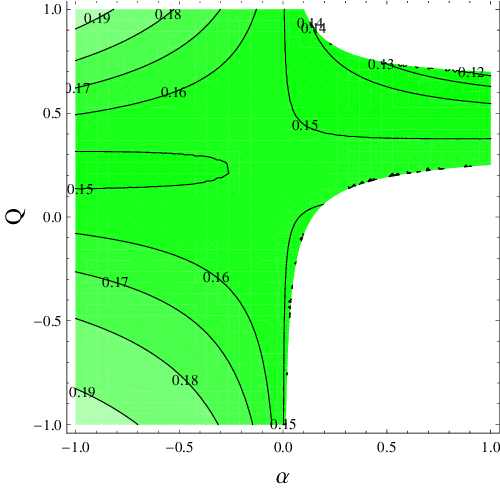} \includegraphics{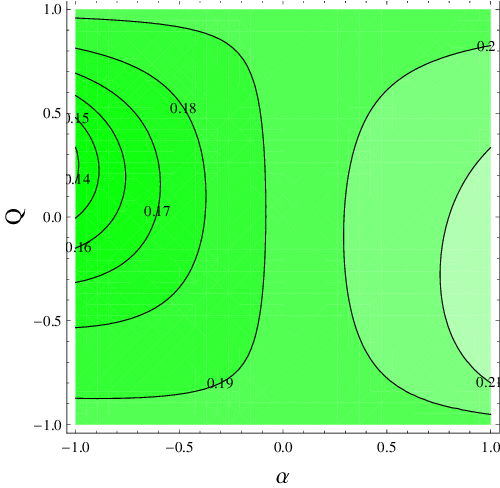}}
\caption{The Lyapunov
 exponent for $\beta=0.1$ (\textbf{left}), $\beta=0.9$ (\textbf{middle}) and $\alpha+\beta=0$ (\textbf{right}), $M=1$.}
\label{fig::Lyapunov}
\end{figure}

\begin{figure}[H]
\centering
\resizebox{0.98\linewidth}{!}{\includegraphics{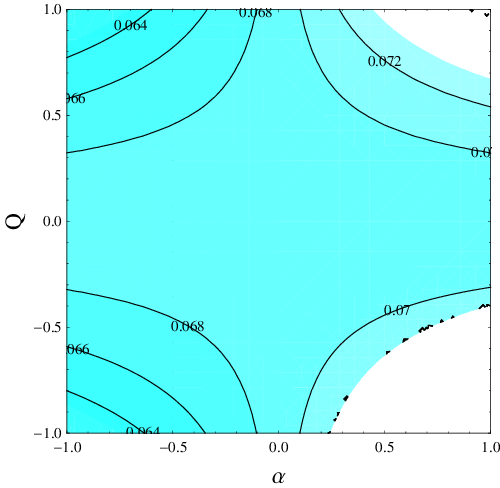}~~\includegraphics{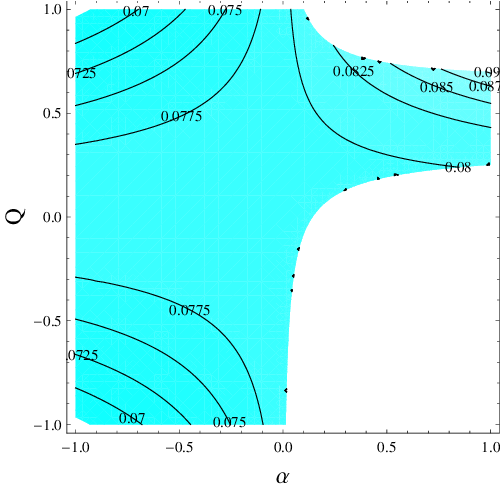}~~\includegraphics{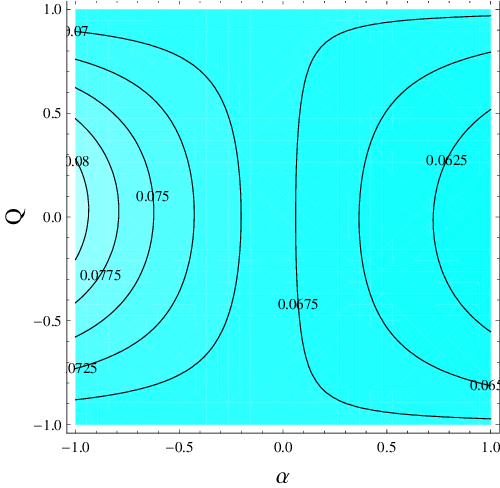}}
\caption{The frequency at ISCO for $\beta=0.1$ (\textbf{left}),  $\beta=0.9$ (\textbf{middle})  and $\alpha+\beta=0$ (\textbf{right}), $M=1$.}
\label{fig:ISCO}
\end{figure}

The radius of the shadow, the Lyapunov exponent, and the ISCO frequency for various parameter values are shown in Figures~\ref{fig:Shadow},~\ref{fig::Lyapunov}, and~\ref{fig:ISCO}, respectively. Due to the established correspondence between null geodesics and QNMs in the eikonal regime~\cite{Cardoso:2008bp}, both the Lyapunov exponent and the shadow radius are directly related to the damping rate of eikonal QNMs~\cite{Cardoso:2008bp,Jusufi:2020dhz} for spherically symmetric and asymptotically flat black holes, with some notable exceptions discussed in~\cite{Konoplya:2022gjp,Bolokhov:2023dxq,Bolokhov:2025uxz}. While the literature on the shadow, Lyapunov exponent, and ISCO characteristics is extensive (see, for example,~\cite{Harada:2010yv,Mondal:2021exj,Schroven:2020ltb,Perlick:2021aok,Perlick:2015vta,Jefremov:2015gza,Konoplya:2021slg,Stuchlik:2019uvf,Tsukamoto:2014tja}), the binding energy has received comparatively less attention~\cite{Esteban:1988zp,Konoplya:2018arm}. Some of these works also provide a general formalism for computing the aforementioned quantities.

\vspace{6pt}
\vspace{-20pt}

\vspace{-20pt}

We shall use the auxiliary function \cite{Konoplya:2020hyk}:
\begin{equation}\label{auxiliary}
P(r )\equiv \frac{f(r )}{r ^2},
\end{equation}
allowing one to identify the {\it radiation zone}, which is conditionally interpreted as a region where classical radiation processes are significant. The latter implies that the near-horizon layer and the asymptotic zone (sufficiently far from the ISCO) are excluded from the radiation zone.

Defining the four-momentum of a particle of mass $m$ as
\begin{equation}\label{momentum}
p^{\mu}\equiv m\frac{dx^{\mu}}{ds},
\end{equation}
where $s$ is an invariant affine parameter, and introducing the energy $E \equiv -p_t$ and angular momentum $L\equiv p_{\phi}$, from the normalization condition on the four-momentum we have
\begin{equation}\label{normalization}
p_{\mu}p^{\mu}=-m^2.
\end{equation}

Due to spherical symmetry, without loss of generality we consider the motion in the equatorial plane $\theta=\pi/2$, $d\theta=0$ and obtain the following equation for the radial coordinate
\begin{equation}\label{radialeq}
m^2 \frac{1}{f(r )}\left(\frac{d r }{ds}\right)^2=V_{eff},
\end{equation}
where the effective potential is determined as
\begin{eqnarray}\label{potential}
V_{eff}(r )=\frac{E^2}{f(r )}-\frac{L^2}{r ^2}-m^2
\end{eqnarray}

The circular motion corresponds to the constant radial coordinate, so that $dr =0$ and $d^2r =0$, which is equivalent to
\begin{equation}\label{circulareq}
\begin{array}{rcl}
V_{eff}(r ) &=& 0, \\
V_{eff}'(r ) &=& 0.
\end{array}
\end{equation}

Then, $E$ and $L$ at the given circular orbit $r $ are
\begin{equation}\label{circularEL}
\begin{array}{rcl}
E^2&=&-m^2\dfrac{2r P^2(r )}{P'(r )},\\
L^2&=&-m^2\dfrac{f'(r )}{P'(r )},\\
\end{array}
\end{equation}
and the corresponding frequency of rotation is defined as
\begin{equation}\label{frequency}
\Omega^2\equiv\left(\frac{d\phi}{dt}\right)^2=\frac{L^2f^2(r)}{E^2r ^4}=\frac{f'(r )}{2r }.
\end{equation}

For the massless particle \mbox{($m=0$)}, circular photon orbits exist only at a single radius, corresponding to the maximum of the auxiliary function $P(r)$. The shadow radius is defined as the impact parameter of photons that escape from the vicinity of the circular photon orbit, $R_s=L/E$. Therefore, using (\ref{circularEL}), $R_s$ can be determined by finding the maximum value of $P(r )$.
\begin{equation}\label{shadow}
\frac{1}{R_{s}^2}=\frac{E^2}{L^2}=\max P(r )=P(r _m).
\end{equation}

Here, $r_m $ denotes the radius at which the auxiliary function attains its maximum, satisfying $P'(r _m)=0$.


After substituting, $$ds^2=\frac{m^2}{E^2}f(r )^2dt^2$$ into (\ref{radialeq}), taking the limit \mbox{$m\to0$}, and considering a deviation from the circular orbit $r $, $r =r _m+\delta r $, one finds the equation for the radial coordinate of such photons,
\begin{equation}\label{LypunovEq}
\left(\frac{d}{dt}\delta r \right)^2=\lambda^2\delta r ^2+\mathcal{O}({\delta r })^3,
\end{equation}
where $\lambda$ is the Lyapunov exponent, which can be calculated using the following relation:
\begin{equation}\label{Lyapunov}
\lambda^2 = -\frac{1}{P}\frac{d^2P}{dr_*^2}\Biggr|_{r =r _m},
\end{equation}
where the tortoise coordinate $r_*$ is defined as
\begin{equation}\label{tortoise_alt}
\frac{dr_*}{dr} = \frac{1}{f(r)}.
\end{equation}

Similarly, for a massive particle at a circular orbit $r$, the radial perturbation equation becomes:
\begin{equation}\label{radialmass}
m^2 \frac{1}{f(r )}\left(\frac{d\delta}{ds}r \right)^2=\frac{V_{eff}''(r )}{2}\delta r ^2+\mathcal{O}({\delta r })^3.
\end{equation}

Therefore, the orbit is stable for all $r$, such that
\begin{equation}\label{stableorbitscond}
V_{eff}''(r )=-4m^2\frac{1}{r }\left(\frac{d\ln E}{dr }\right)<0.
\end{equation}

Note that here the value of the second derivative of $V_{eff}$ for constant $E$ and $L$, satisfying the condition (\ref{circularEL}), is given as a function of the circular orbit coordinate value $r $.

In the asymptotically flat space-time, the marginally stable circular orbit, i.e., the one for which,
\begin{equation}\label{MSCO}
V_{eff}''(r )=0,
\end{equation}
corresponds to the ISCO with the minimum of energy,
\begin{equation}\label{EISCO}
E_{ISCO}^2=-m^2\max\dfrac{2r  P^2(r )}{P'(r )},
\end{equation}
which is attained at the orbit of $r =r _{ISCO}$,
\begin{equation}\label{ISCO}
V_{eff}''(r _{ISCO})=0.
\end{equation}

The invariant characteristics of the ISCO are its frequency $\Omega_{ISCO}$, calculated using (\ref{frequency}) for $r =r _{ISCO}$ and the binding energy, i.e., the amount of energy per unit mass released by a particle going over from the distant orbit with $E\simeq m$,
\begin{equation}\label{BEdef}
BE=\frac{E-E_{ISCO}}{m}=1-\frac{E_{ISCO}}{m}.
\end{equation}

Tables~\ref{tab:1}--\ref{tab:4} present the values of all four characteristics for the metric (\ref{metricfunc}), evaluated at various fixed values of the parameters $\alpha$, $\beta$, and $Q$, with the black hole mass fixed at $M = 1$. These results show that, unless $\alpha$ and $Q$ exceed unity, their influence on the observables is relatively mild, with deviations from the Schwarzschild limit remaining within a few percent.

The corresponding quantities for different values of $Q$ and $\beta = -\alpha $ are shown in Table~\ref{tab:5}. As $\alpha$ increases from negative values through zero to positive values, we observe that the shadow radius and the Lyapunov exponent increase monotonically, whereas the ISCO frequency and the binding energy decrease.

\begin{table}[H]
\fontsize{9}{9}\selectfont 
\caption{Shadow radius, Lyapunov exponent, ISCO frequency and binding energy for $M=1$, $\beta=0.1$.}
\newcolumntype{c}{>{\centering\arraybackslash}X}

\begin{tabularx}{\textwidth}{cccccc}
\toprule
\boldmath{$\alpha$} & \boldmath{$Q$} & \boldmath{$R_{sh}$} & \boldmath{$\lambda$} & \boldmath{$\Omega_{ISCO}$} & \textbf{BE}\\
\midrule
$-1.0$ & $-0.5$ & $5.24717$ & $0.195170$ & $0.06678$ & $0.05657$\\
$-0.7$ & $-0.5$ & $5.22266$ & $0.194002$ & $0.06739$ & $0.05687$\\
$-0.5$ & $-0.5$ & $5.20523$ & $0.193019$ & $0.06782$ & $0.05708$\\
$-0.2$ & $-0.5$ & $5.17716$ & $0.191125$ & $0.06850$ & $0.05741$\\
$0.2$ & $-0.5$ & $5.13522$ & $0.187434$ & $0.06949$ & $0.05787$\\
$0.5$ & $-0.5$ & $5.09923$ & $0.183198$ & $0.07029$ & $0.05824$\\
$0.7$ & $-0.5$ & $5.07230$ & $0.179164$ & $0.07087$ & $0.05850$\\
\midrule
$-1.0$ & $0.5$ & $5.25436$ & $0.197108$ & $0.06671$ & $0.05655$\\
$-0.7$ & $0.5$ & $5.22536$ & $0.194751$ & $0.06736$ & $0.05687$\\
$-0.5$ & $0.5$ & $5.20591$ & $0.193210$ & $0.06781$ & $0.05708$\\
$-0.2$ & $0.5$ & $5.17658$ & $0.190953$ & $0.06851$ & $0.05741$\\
$0.2$ & $0.5$ & $5.13720$ & $0.188059$ & $0.06947$ & $0.05787$\\
$0.5$ & $0.5$ & $5.10749$ & $0.185990$ & $0.07022$ & $0.05822$\\
$0.7$ & $0.5$ & $5.08762$ & $0.184665$ & $0.07073$ & $0.05846$\\
\bottomrule
\end{tabularx}
\label{tab:1}
\end{table}
\vspace{-20pt}

\begin{table}[H]
\fontsize{9}{9}\selectfont 
\caption{Shadow radius, Lyapunov exponent, ISCO frequency and binding energy for $M=1$, $\beta=0.3$.}
\newcolumntype{c}{>{\centering\arraybackslash}X}
\begin{tabularx}{\textwidth}{cccccc}
\toprule
\boldmath{$\alpha$} & \boldmath{$Q$} & \boldmath{$R_{sh}$} & \boldmath{$\lambda$} & \boldmath{$\Omega_{ISCO}$} & \textbf{BE}\\
\midrule
$-1.0$ & $-0.5$ & $5.17872$ & $0.191530$ & $0.06848$ & $0.05741$\\
$-0.7$ & $-0.5$ & $5.15071$ & $0.189784$ & $0.06919$ & $0.05774$\\
$-0.5$ & $-0.5$ & $5.13057$ & $0.188299$ & $0.06968$ & $0.05797$\\
$-0.2$ & $-0.5$ & $5.09766$ & $0.185383$ & $0.07047$ & $0.05834$\\
$0.2$ & $-0.5$ & $5.04713$ & $0.179392$ & $0.07163$ & $0.05886$\\
$0.3$ & $-0.5$ & $5.03291$ & $0.177283$ & $0.07194$ & $0.05900$\\
\midrule
$-1.0$ & $0.5$ & $5.17596$ & $0.190758$ & $0.06851$ & $0.05741$\\
$-0.7$ & $0.5$ & $5.14547$ & $0.188272$ & $0.06924$ & $0.05776$\\
$-0.5$ & $0.5$ & $5.12502$ & $0.186655$ & $0.06974$ & $0.05799$\\
$-0.2$ & $0.5$ & $5.09418$ & $0.184299$ & $0.07051$ & $0.05835$\\
$0.2$ & $0.5$ & $5.05278$ & $0.181311$ & $0.07158$ & $0.05885$\\
$0.3$ & $0.5$ & $5.04239$ & $0.180595$ & $0.07186$ & $0.05898$\\
\bottomrule
\end{tabularx}
\label{tab:2}
\end{table}
\vspace{-20pt}

\begin{table}[H]
\fontsize{9}{9}\selectfont 
\caption{Shadow radius, Lyapunov exponent, ISCO frequency and binding energy for $M=1$, $\beta=0.5$.}
\newcolumntype{c}{>{\centering\arraybackslash}X}
\begin{tabularx}{\textwidth}{cccccc}
\toprule
\boldmath{$\alpha$} & \boldmath{$Q$} & \boldmath{$R_{sh}$} & \boldmath{$\lambda$} & \boldmath{$\Omega_{ISCO}$} & \textbf{BE}\\
\midrule
$0.01$ & $0.3$ & $4.98163$ & $0.174659$ & $0.07339$ & $0.05967$\\
$0.1$ & $0.3$ & $4.97853$ & $0.174545$ & $0.07348$ & $0.05971$\\
$0.3$ & $0.3$ & $4.97169$ & $0.174308$ & $0.07369$ & $0.05981$\\
$0.5$ & $0.3$ & $4.96491$ & $0.174092$ & $0.07390$ & $0.05991$\\
$0.7$ & $0.3$ & $4.95820$ & $0.173897$ & $0.07412$ & $0.06000$\\
$0.9$ & $0.3$ & $4.95155$ & $0.173722$ & $0.07433$ & $0.06010$\\
$0.9$ & $0.3$ & $4.95155$ & $0.173722$ & $0.07433$ & $0.06010$\\
$0.95$ & $0.3$ & $4.94989$ & $0.173681$ & $0.07438$ & $0.06013$\\
\midrule
$0.01$ & $0.7$ & $4.97971$ & $0.174482$ & $0.07344$ & $0.05969$\\
$0.1$ & $0.7$ & $4.95910$ & $0.172752$ & $0.07400$ & $0.05994$\\
$0.3$ & $0.7$ & $4.91215$ & $0.168815$ & $0.07529$ & $0.06052$\\
$0.5$ & $0.7$ & $4.86353$ & $0.164757$ & $0.07669$ & $0.06113$\\
$0.7$ & $0.7$ & $4.81309$ & $0.160592$ & $0.07818$ & $0.06178$\\
$0.9$ & $0.7$ & $4.76070$ & $0.156341$ & $0.07981$ & $0.06247$\\
$0.9$ & $0.7$ & $4.76070$ & $0.156341$ & $0.07981$ & $0.06247$\\
$0.95$ & $0.7$ & $4.74728$ & $0.155269$ & $0.08023$ & $0.06265$\\
\bottomrule
\end{tabularx}
\label{tab:3}
\end{table}
\vspace{-20pt}

\begin{table}[H]
\fontsize{9}{9}\selectfont 
\caption{Shadow radius, Lyapunov exponent, ISCO frequency and binding energy for $M=1$, $\beta=0.7$.}
\newcolumntype{c}{>{\centering\arraybackslash}X}
\begin{tabularx}{\textwidth}{cccccc}
\toprule
\boldmath{$\alpha$} & \boldmath{$Q$} & \boldmath{$R_{sh}$} & \boldmath{$\lambda$} & \boldmath{$\Omega_{ISCO}$} & \textbf{BE}\\
\midrule
$0.01$ & $0.3$ & $4.87925$ & $0.164419$ & $0.07611$ & $0.06087$\\
$0.1$ & $0.3$ & $4.87621$ & $0.164390$ & $0.07622$ & $0.06091$\\
$0.3$ & $0.3$ & $4.86951$ & $0.164342$ & $0.07645$ & $0.06102$\\
$0.5$ & $0.3$ & $4.86289$ & $0.164314$ & $0.07669$ & $0.06113$\\
$0.7$ & $0.3$ & $4.85635$ & $0.164306$ & $0.07692$ & $0.06124$\\
$0.9$ & $0.3$ & $4.84988$ & $0.164317$ & $0.07716$ & $0.06135$\\
$0.9$ & $0.3$ & $4.84988$ & $0.164317$ & $0.07716$ & $0.06135$\\
$0.95$ & $0.3$ & $4.84827$ & $0.164322$ & $0.07722$ & $0.06137$\\
\midrule
$0.01$ & $0.7$ & $4.87708$ & $0.164193$ & $0.07617$ & $0.06089$\\
$0.1$ & $0.7$ & $4.85423$ & $0.162101$ & $0.07682$ & $0.06117$\\
$0.3$ & $0.7$ & $4.80190$ & $0.157302$ & $0.07834$ & $0.06183$\\
$0.5$ & $0.7$ & $4.74725$ & $0.152300$ & $0.08000$ & $0.06253$\\
$0.7$ & $0.7$ & $4.69004$ & $0.147114$ & $0.08180$ & $0.06328$\\
$0.9$ & $0.7$ & $4.63001$ & $0.141783$ & $0.08379$ & $0.06409$\\
$0.9$ & $0.7$ & $4.63001$ & $0.141783$ & $0.08379$ & $0.06409$\\
$0.95$ & $0.7$ & $4.61453$ & $0.140436$ & $0.08432$ & $0.06430$\\
\bottomrule
\end{tabularx}
\label{tab:4}
\end{table}
\vspace{-20pt}

\begin{table}[H]
\fontsize{9}{9}\selectfont 
\caption{Shadow radius, Lyapunov exponent, ISCO frequency and binding energy for $\beta=-\alpha$, $M=1$.}
\newcolumntype{c}{>{\centering\arraybackslash}X}
\begin{tabularx}{\textwidth}{cccccc}
\toprule
\boldmath{$\alpha$} & \boldmath{$Q$} & \boldmath{$R_{sh}$} & \boldmath{$\lambda$} & \boldmath{$\Omega_{ISCO}$} & \textbf{BE}\\
\midrule
$-1.0$ & $-0.8$ & $5.07817$ & $0.188300$ & $0.07135$ & $0.05880$\\
$-0.5$ & $-0.8$ & $5.13205$ & $0.189030$ & $0.06969$ & $0.05799$\\
$-0.1$ & $-0.8$ & $5.18250$ & $0.191497$ & $0.06837$ & $0.05735$\\
$0.1$ & $-0.8$ & $5.21021$ & $0.193551$ & $0.06771$ & $0.05703$\\
$0.5$ & $-0.8$ & $5.27028$ & $0.199470$ & $0.06641$ & $0.05642$\\
$1.0$ 
 & $-0.8$ & $5.35267$ & $0.209978$ & $0.06483$ & $0.05567$\\
\midrule
$-1.0$ & $-0.3$ & $4.80843$ & $0.169131$ & $0.07899$ & $0.06217$\\
$-0.5$ & $-0.3$ & $5.01115$ & $0.179242$ & $0.07275$ & $0.05939$\\
$-0.1$ & $-0.3$ & $5.16074$ & $0.189831$ & $0.06889$ & $0.05760$\\
$0.1$ & $-0.3$ & $5.23079$ & $0.195015$ & $0.06723$ & $0.05680$\\
$0.5$ & $-0.3$ & $5.36203$ & $0.204552$ & $0.06431$ & $0.05537$\\
$1.0$ & $-0.3$ & $5.51141$ & $0.214682$ & $0.06124$ & $0.05383$\\
\midrule
$-1.0$ & $0.3$ & $4.72590$ & $0.138767$ & $0.07979$ & $0.06238$\\
$-0.5$ & $0.3$ & $4.99945$ & $0.175394$ & $0.07286$ & $0.05942$\\
$-0.1$ & $0.3$ & $5.16040$ & $0.189729$ & $0.06890$ & $0.05760$\\
$0.1$ & $0.3$ & $5.23050$ & $0.194931$ & $0.06723$ & $0.05680$\\
$0.5$ & $0.3$ & $5.35638$ & $0.203035$ & $0.06436$ & $0.05539$\\
$1.0$ & $0.3$ & $5.49405$ & $0.210415$ & $0.06139$ & $0.05388$\\
\midrule
$-1.0$ & $0.8$ & $5.04241$ & $0.179002$ & $0.07175$ & $0.05892$\\
$-0.5$ & $0.8$ & $5.12345$ & $0.186645$ & $0.06978$ & $0.05801$\\
$-0.1$ & $0.8$ & $5.18217$ & $0.191400$ & $0.06837$ & $0.05735$\\
$0.1$ & $0.8$ & $5.20989$ & $0.193454$ & $0.06772$ & $0.05704$\\
$0.5$ & $0.8$ & $5.26254$ & $0.197063$ & $0.06648$ & $0.05644$\\
$1.0$ & $0.8$ & $5.32392$ & $0.200843$ & $0.06507$ & $0.05574$\\
\bottomrule
\end{tabularx}
\label{tab:5}
\end{table}

\section{Grey-Body Factors}\label{sec:GBF}

\subsection{Perturbation Equations}

The dynamics of scalar and Dirac fields in curved spacetime are governed by covariant field equations that incorporate the geometry of the background. In a general relativistic setting, these equations for the scalar field $\Phi$ and the spin-$\tfrac{1}{2}$ Dirac field $\Upsilon$ take the form:
\begin{equation}\label{coveqs}
\begin{array}{rcl}
\partial_\mu \left(\sqrt{-g} \, g^{\mu \nu} \, \partial_\nu \Phi \right) &=& 0, \\
\gamma^{\alpha} \left( \partial_\alpha - \Gamma_{\alpha} \right) \Upsilon &=& 0,
\end{array}
\end{equation}
where $\gamma^{\alpha}$ are gamma matrices defined in the local frame, and $\Gamma_{\alpha}$ denote the associated spin connections as defined in the tetrad formalism. The scalar field equation corresponds to the covariant Klein--Gordon equation, while the Dirac equation follows from its curved-spacetime generalization.

Upon separation of variables in a static, spherically symmetric black hole background, both equations reduce to a Schrödinger-like master equation of the form:
\begin{equation}\label{master_wave}
\frac{d^2 \Psi}{dr_*^2} + \left( \omega^2 - V(r) \right) \Psi = 0,
\end{equation}
to regularize the behavior near the horizon and map the infinite domain into $\mathbb{R}$.

The effective potential governing scalar perturbations is given by:
\begin{equation}\label{V_scalar_alt}
V(r) = \frac{\ell(\ell+1)}{r^2} f(r) + \frac{1}{r} \frac{d^2 r}{dr_*^2},
\end{equation}
where $\ell = 0, 1, 2, \cdots$ is the multipole number.

For massless Dirac fields, the perturbations are reduced to two wavelike equations with the following effective potentials:
\begin{equation}\label{V_dirac_alt}
V_{\pm}(r) = W^2 \pm \frac{dW}{dr_*}, \quad W = \frac{\sqrt{f(r)}}{r} \left( \ell + \tfrac{1}{2} \right),
\end{equation}
where the multipole numbers take half-integer values \,\, $\ell = 1/2, 3/2, 5/2, \cdots$ .

These two potentials, $V_+$ and $V_-$, are isospectral, and the corresponding solutions are related through a Darboux transformation:
\begin{equation}\label{darboux}
\Psi_{+} \propto \left( W + \frac{d}{dr_*} \right) \Psi_{-},
\end{equation}
which implies that it is sufficient to compute physical quantities such as GBFs using only one of the two potentials, typically $V_+(r)$.

\subsection{Boundary Conditions and Definition}

In asymptotically flat spacetimes, the appropriate boundary conditions for computing the GBFs correspond to purely ingoing waves at the event horizon and a superposition of ingoing and outgoing waves at spatial infinity:
\begin{align}
\Psi(r_*) &\sim e^{-i \omega r_*}, \quad r_* \to -\infty \quad \text{(horizon)}, \\
\Psi(r_*) &\sim A_{\text{in}} e^{-i \omega r_*} + A_{\text{out}} e^{i \omega r_*}, \quad r_* \to +\infty.
\end{align}
The transmission coefficient \( \mathcal{T}_\ell(\omega) \) for a given multipole number \( \ell \) is then defined as
\begin{equation}
|\mathcal{T}_\ell(\omega)|^2 = \left| \frac{A_{\text{trans}}}{A_{\text{in}}} \right|^2 = 1 - \left| \frac{A_{\text{out}}}{A_{\text{in}}} \right|^2,
\end{equation}
where \( A_{\text{trans}} \) is the amplitude of the wave transmitted through the potential barrier. The quantity $\Gamma(\Omega) = |\mathcal{T}_\ell(\omega)|^2 $ serves as the GBF.

\subsection{Higher-Order WKB Method and the Correspondence Between GBFs and QNMs}

An efficient semi-analytical approach to calculate the GBFs is the WKB method, which is particularly useful for frequencies near the peak of the effective potential. The WKB method for QNMs has been formulated in \cite{Schutz:1985km} at the first WKB order and developed to higher orders in \cite{Iyer:1986np,Konoplya:2003ii,Matyjasek:2017psv}. Its \( N \)th-order implementation approximates the transmission coefficient as
\begin{equation}
|\mathcal{T}_\ell(\omega)|^2 = \left(1 + e^{2 i \pi K(\omega)} \right)^{-1},
\end{equation}
where the function \( K(\omega) \) depends on the properties of the potential at its maximum and includes higher-order corrections:
\begin{equation}
K(\omega) = \frac{i (\omega^2 - V_0)}{\sqrt{-2 V_0''}} + \sum_{j=2}^{N} \Lambda_j,
\end{equation}
with \( V_0 \equiv V(r_0) \) being the value of the potential at its peak \( r_0 \), \( V_0'' \equiv d^2V/dr_*^2|_{r_*=r_0} \), and \( \Lambda_j \) denoting higher-order correction terms, whose explicit forms are known up to the 13th order \cite{Matyjasek:2017psv}. However, we will use the sixth order formula \cite{Konoplya:2003ii}, which is known to be optimal in the majority of cases. The method becomes increasingly accurate for higher \( \ell \).

Alternatively, we compute the GBFs using their correspondence with QNMs~\cite{Lutfuoglu:future}. Specifically, we relate the grey-body factor $\Gamma_\ell(\Omega)$ to the least damped quasinormal mode $\omega_0$ via the expression~\cite{Konoplya:2024lir}
\[
\Gamma_\ell(\Omega)  = \left(1 + \exp\left[\frac{2\pi\left(\Omega^2 - \text{Re}(\omega_0)^2\right)}{4\, \text{Re}(\omega_0)\, |\text{Im}(\omega_0)|} \right]\right)^{-1} + \mathcal{O}(\ell^{-1}).
\]

This correspondence has recently been explored in~\cite{Malik:2024cgb,Bolokhov:2024otn,Pedrotti:2025upg,Hamil:2025cms,Lutfuoglu:2025ljm,Dubinsky:2024vbn,Skvortsova:2024msa,Tang:2025mkk,Lutfuoglu:2025ohb}. The correction terms beyond the eikonal approximation, not shown explicitly here, depend on the first overtone of the quasinormal spectrum. In our computations, we incorporate these corrections up to third order beyond the leading eikonal limit.

\begin{figure}[H]
\centering
\resizebox{0.9\linewidth}{!}{\includegraphics{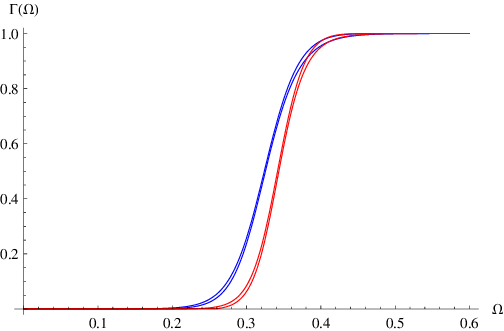}\includegraphics{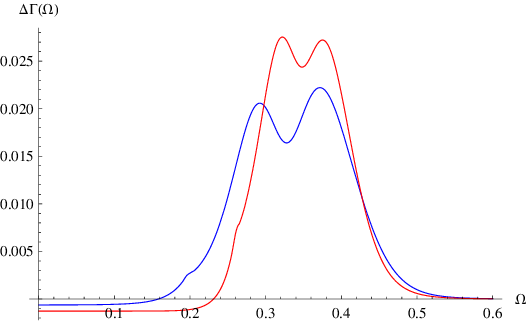}}
\caption{\textls[-15]{GBFs (left) and the absolute difference between GBFs calculated via the 6th order WKB method and the correspondence with QNMs. Scalar field perturbations: $\ell=1$, $\alpha=0.6$, $\beta=1$, $Q=0.05$ (blue) and $Q=0.75$ (red).}}
\label{fig:GBFL1}
\end{figure}

\begin{figure}[H]
\centering
\resizebox{0.9\linewidth}{!}{\includegraphics{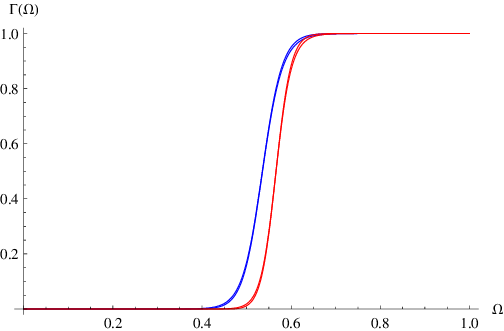}\includegraphics{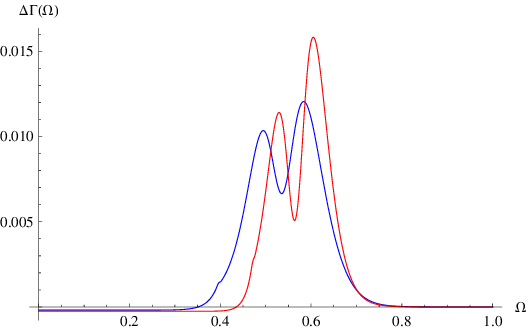}}
\caption{\textls[-15]{GBFs (left) and the absolute difference between GBFs calculated via the 6th order WKB method and the correspondence with QNMs. Scalar field perturbations: $\ell=2$, $\alpha=0.6$, $\beta=1$, $Q=0.05$ (blue) and $Q=0.75$ (red).}}
\label{fig:GBFL2}
\end{figure}

\begin{figure}[H]
\centering
\resizebox{0.9\linewidth}{!}{\includegraphics{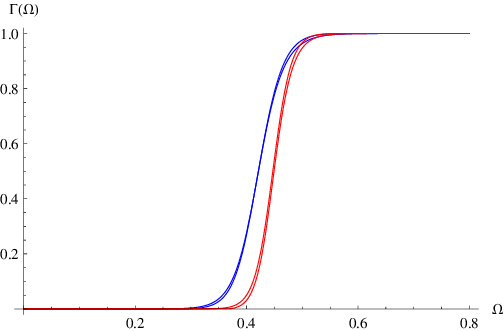}\includegraphics{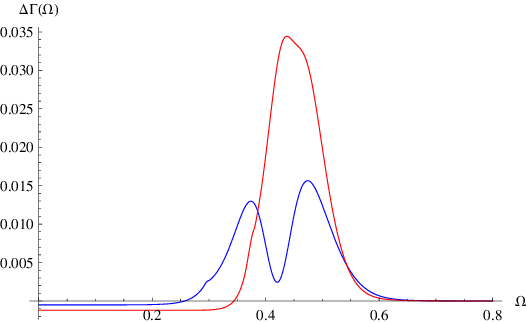}}
\caption{\textls[-15]{GBFs (left) and the absolute difference between GBFs calculated via the 6th order WKB method and the correspondence with QNMs. Dirac field perturbations: $\ell=3/2$, $\alpha=0.6$, $\beta=1$, $Q=0.05$ (blue) and $Q=0.75$ (red).}}
\label{fig:GBFL3over2}
\end{figure}

As illustrated in Figures~\ref{fig:GBFL1}--\ref{fig:GBFL3over2}, the correspondence yields reliable results only for multipole numbers $\ell > 1$ in the case of scalar fields and $\ell > \tfrac{1}{2}$ for Dirac perturbations. For scalar modes with $\ell = 1$, the relative error exceeds 10\%, while for $\ell = 0$, the discrepancy can reach several tens of percent.
We also observe that the GBFs decrease as the parameter $Q$ increases. This behavior can be attributed to the shape of the effective potential, which becomes higher for larger values of $Q$, thereby suppressing the transmission probability.

\vspace{-20pt}

\vspace{-20pt}

\vspace{-22pt}

\section{Conclusions}\label{sec:conclusion}

In this work, we have analyzed the family of asymptotically flat black hole solutions in a recently proposed model of Proca–Gauss–Bonnet gravity, characterized by vector–tensor and scalar–tensor higher-curvature couplings. These solutions feature nontrivial primary hair and yield two distinct metric forms depending on the relation between the coupling constants. For each case, we determined the parametric regions that allow for the existence of an event horizon and explored both classical and semiclassical observables.

On the classical side, we studied particle motion and photon trajectories, focusing on the shadow radius, Lyapunov exponent, ISCO frequency, and binding energy. Our results indicate that for moderate values of the couplings, the deviations from the Schwarzschild limit remain small, but become significant when the couplings or hair parameter grow. The second metric, obtained under the condition $\alpha = -\beta$, shows a monotonic behavior in several observables, making it a useful testbed for probing deviations from general relativity.

On the semiclassical side, we analyzed scalar and Dirac perturbations and evaluated the corresponding GBFs using both the sixth-order WKB method and the recently proposed correspondence with QNMs. We found that this correspondence offers a reliable estimate of the GBFs for multipole numbers $\ell > 1$ for scalar fields and $\ell > 1/2$ for Dirac fields. The accuracy degrades significantly for lower multipoles. We also observed that increasing the Proca hair parameter $Q$ leads to a higher effective potential, thereby reducing the transmission probability. The obtained grey-body factors can be further used to calculate the energy and charge emission rates of black holes, providing estimates for the intensity of Hawking evaporation.

From observations of shadows cast by black holes \cite{EventHorizonTelescope:2019dse,EventHorizonTelescope:2019ggy} we have \cite{Vagnozzi:2022moj}
\begin{equation}
4.55M \lesssim R_{\text{sh}} \lesssim 5.22M, \quad (1\sigma).
\end{equation}

Using the above relation, one can place constraints on other parameters of the black hole, such as the charge $Q$ and the couplings $\alpha$ and $\beta$. Based on the data presented in this work, it is evident that the constraints on $\alpha$, $\beta$, and $Q$ are very weak, allowing for significant deviations from the Schwarzschild geometry. Apparently, accounting for more subtle optical phenomena related to radiation in the presence of an accretion disk could lead to tighter constraints; however, these would strongly depend on the equation of state of the matter surrounding the black hole.

Altogether, our study demonstrates that the presence of primary hair and higher-curvature corrections has a measurable impact on both classical trajectories and grey-body radiation spectra, potentially offering new avenues for testing modified gravity theories with future astrophysical observations.

Our work could be naturally extended to the analysis of quasi-periodic oscillations (QPOs) in the vicinity of the considered black holes, offering potential observational signatures of the underlying geometry.

In future studies, it would be interesting to investigate how the parameters of the black hole, such as charge and coupling constants, affect the frequencies of quasi-period oscillations (QPOs) and compare the predictions with observational data. Another promising direction is to explore the impact of the effective potential on the epicyclic frequencies of particle motion, which are closely related to the origin of QPOs in accretion disks. Such an extension may provide a link between theoretical models and astrophysical observations, particularly in the context of high-frequency QPOs detected in X-ray binaries \cite{Bambi:2015kza}.
	


		\section*{Funding}
The author is grateful to Excellence Project PrF UHK 2205/2025-2026.
 
\section*{Institutional Review Board Statement}
Not applicable.

\section*{Informed Consent Statement}
Not applicable.

\section*{Data Availability Statement}
Not applicable.

		\section*{Conflicts of Interest}
The authors declare no conflict of interest.

	\small
	\bibliographystyle{scilight}
	
	

\end{document}